\documentclass[amssymb, nobibnotes, aps, pra,floatfix,superscriptaddress,notitlepage,twocolumn]{revtex4-1}

\usepackage{graphicx} 
\usepackage{hyperref}
\begin{document}

\title{Mobility and Clustering of Barium Ions and Dications in High Pressure Xenon Gas}

\author{E. Bainglass}
\thanks{Corresponding author: \url{edan.bainglass@mavs.uta.edu}}
\affiliation{Department of Physics, University of Texas at Arlington,
Arlington, Texas 76019, USA}
\author{B.J.P. Jones}
\thanks{Corresponding author: \url{ben.jones@uta.edu}}
\affiliation{Department of Physics, University of Texas at Arlington,
Arlington, Texas 76019, USA}
\author{F. W. Foss Jr}
\affiliation{Department of Chemistry, University of Texas at Arlington,
Arlington, Texas 76019, USA}
\author{M. N. Huda}
\affiliation{Department of Physics, University of Texas at Arlington,
Arlington, Texas 76019, USA}
\author{D. R. Nygren}
\affiliation{Department of Physics, University of Texas at Arlington,
Arlington, Texas 76019, USA}


\begin{abstract}
The clustering and drift properties of barium ions in xenon gas are explored theoretically, using density functional theory and computational ion mobility theory, with the goal of better understanding barium ion transport for neutrinoless double beta decay.  We derive the equilibrium conformations, energies and entropies of molecular ions in the Ba$^{+}$-Xe and Ba$^{++}$-Xe systems, which yield a predictive model of cluster formation in high pressure gas.  We calculate ion-neutral interaction potential curves for these species and use them to predict effective molecular ion mobilities.  Our calculation consistently reproduces experimental data on effective mobility and molecular ion formation for the Ba$^+$ system, and predicts strong cluster formation in the Ba$^{++}$ system, dominated by stable [BaXe$_6$]$^{++}$,[BaXe$_7$]$^{++}$, [BaXe$_8$]$^{++}$ and [BaXe$_9$]$^{++}$ complexes in the range of interest.  Some implications for barium tagging in gas-phase neutrinoless double beta decay experiments are discussed, and the first predictions of pressure-dependent mobility of the doubly charged Ba$^{++}$ species are presented.

\end{abstract}
 
\maketitle
\flushbottom

\section{Introduction and Motivation \label{sec:Intro}}

The production and detection of barium ions in a gaseous xenon environment is a topic of particular interest in experimental nuclear physics. This is because the detection of such ions could serve as an unambiguous tag of the hypothetical, ultra-rare process of neutrinoless double beta decay \cite{Moe:1991ik}.  So-called ``barium tagging'' has been a subject of intensive R\&D for both liquid and gaseous xenon detectors \cite{McDonald:2017izm,Sinclair:2011zz,Mong:2014iya}.  However, only recently has single ion sensitivity been demonstrated using a method that appears compatible with {\em{in-situ}}  ion identification \cite{Jones:2016qiq}.  To realize barium detection through ion drift followed by fluorescence sensing in high pressure xenon gas experiments, it is necessary to understand the mobility of barium ions in the pressure and electric field ranges of interest for time projection chamber detector (TPC) experiments \cite{Martin-Albo:2015rhw}. These are typically in the range 10-15 bar, and 200-500 V/cm, respectively. 

In this paper we present new calculations of the effective mobility of Ba$^{+}$ and Ba$^{++}$ ions in xenon gas.  The main motivation for this work is to produce a mobility prediction that consistently accounts for the effects of molecular ion formation in the barium / xenon system. The association and dissociation of molecular ions $\mathrm{[BaXe_N]}^{q+}$ is a dynamical process that is continuously active in xenon gas at pressures above $\gtrsim$~0.1~bar for Ba$^+$ and $\gtrsim$~0.01~bar for Ba$^{++}$.  The population distribution and absolute mobility of each molecular species are critical ingredients in determining effective ion mobility in this pressure regime.   As such, our prediction is the first to be applicable near or above atmospheric pressure, including the conditions of interest for barium tagging experiments.  Our primary result is a prediction of the effective mobilities for both species in the pressure range 0.1 - 20 bar, shown in Fig.~\ref{fig:MobilityVsPressure}.

Previous theoretical work on Ba$^{+}$ and Ba$^{++}$ drift in xenon gas \cite{McGuirk2009,Viehland2017} treated bare barium ions using coupled cluster theory.  These predictions apply when pressures are sufficiently low that molecular ion formation is negligible, and have a stated accuracy of $\sim$0.05\% in this regime.   Because such techniques are prohibitively computationally expensive when applied to large multi-atomic systems, we opt for a different approach, using Density Functional Theory (DFT) to predict molecular ion configurations and populations, and then computational ion mobility theory to predict individual species mobilities based on scattering interactions during transport.  Our approach is benchmarked against coupled cluster predictions for bare ions, and validated against existing data for the Ba$^{+}$ system, with few-percent level accuracy found in all cases.  This validates the accuracy of the technique, which is then applied to cases where experimental data and past theoretical work are not available.

The data that exist pertaining to mobility of Ba$^{+}$ in xenon gas span the pressure range of 0-1 bar \cite{Cesar2014}.  To our knowledge, no mobility data exists for the Ba$^{++}$ charge state. The latter is of the greatest interest for gas-phase barium tagging, since it is the expected product of the double beta decay process in xenon gas at pressures below 50~bar, where existing and proposed experiments operate. At higher pressures and in the liquid phase, electron-ion recombination is frequent \cite{1997NIMPA.396..360B} and will produce a distribution of charge states between Ba$^{++}$, Ba$^{+}$ and neutral Ba.  This has been demonstrated in the liquid phase using radon daughters \cite{Albert:2015vma}.  In the gas phase, on the other hand, the dication Ba$^{++}$ is the expected outcome, since its ionization potential of 10.00~eV is many times $kT$ lower than that of xenon, 12.13~eV.  The approximately recombinationless behaviour of daughter ions following beta decays in xenon gas has recently been verified experimentally by the NEXT collaboration, with Ref.~\cite{Novella:2018ewv} demonstrating survival of positively charged radon daughters at the $\sim$99\% level.

It has also been established experimentally \cite{Cesar2014} that in the Ba$^{+}$ system at pressures above a few Torr a pressure-dependent reduced mobility is observed, which can be attributed to the effects of molecular ion formation, in particular $\mathrm{[BaXe]}^{+}$.  The distribution and mobility of clusters / molecular ions, rather than bare atomic ions, is therefore critical for understanding the effective mobility in this pressure range. 

Molecular ion formation in the [BaXe]$^{+}$ system proceeds schematically via the reaction:
\begin{equation}
    \mathrm{[BaXe]}^+\leftrightarrow \mathrm{[Ba]^+} + \mathrm{Xe}.
\end{equation}
This process may be assisted by additional spectator xenon atoms, via ter-molecular association and dissociation:
\begin{equation}
    \mathrm{[BaXe]}^{+}+\mathrm{Xe}\leftrightarrow \mathrm{[Ba]^+} + \mathrm{Xe}+\mathrm{Xe}.
\end{equation}

In either case, the equilibrium is similarly forced toward the left(right) at higher(lower) pressures, in accordance with Le Chatelier's principle.  This gives the effective mobility a strong pressure dependence in the 0-1 bar range.  Measurements of this mobility were used to deduce the equilibrium constant for the reaction and the mobilities of the two species empirically in Ref~\cite{Cesar2014}.  Mobility of bare Ba$^{+}$ was found to less than but within experimental systematic uncertainty of prior predictions for atomic ions \cite{McGuirk2009}. The mobility of the [BaXe]$^{+}$ molecular ion was also reported, with no prior theoretical prediction available for comparison. In this work we predict the full pressure dependent mobility in the Ba$^+$ / Xe system including the mobility of bare barium ions, molecular ions and their equilibrium constants, and finding good agreement with data from Ref~\cite{Cesar2014}.

Molecular ion formation processes may be expected to become significant at much lower pressures for Ba$^{++}$, given its higher charge state and deeper potential well.  At higher pressures we may expect the production of larger clusters via successive reactions:
\begin{equation}
    \mathrm{ \mathrm{[BaXe_{N-1}]^{++}}  + \mathrm{Xe}\leftrightarrow \mathrm{[BaXe_N]}^{++}},
\end{equation}
with N$\geq$1. The forward reaction is always enthalpically favorable, since polarization and binding of the previously free xenon atom always yields a lower energy state.  However, this binding is disfavored entropically, with free xenon atoms enjoying a higher entropy state than bound ones.  This competition between entropy and enthalpy determines the equilibrium cluster sizes, which will thus depend on both temperature and pressure in the general case.

Our approach in this paper will be as follows.  First, the molecular conformation, and enthalpy and entropy of binding for each relevant cluster $\mathrm{[BaXE_N]}^{q+}$ in the Ba$^{+}$ and Ba$^{++}$ systems will be calculated using Density Functional Theory (DFT).  The standard entropy and enthalpy thus derived will be used to predict equilibrium coefficients and  cluster population distributions.  The pressure dependence of the cluster populations can be derived using standard activities in the gas phase, and the temperature dependence can be understood through the dependence of the equilibrium constant on the Gibbs energy. 

The potential surfaces for ion interactions with neutral xenon atoms (henceforth referred to as ion-neutral interactions) are then derived for each cluster $\mathrm{[BaXe_N]}^{q+}$ in the Ba$^{+}$ and Ba$^{++}$ systems, based on the simulated molecular ion equilibrium configuration. This allows computation of the momentum-transfer cross section for scattering at thermal energies, which can be used as an input to predict the low-field ion mobility value. We use techniques from \cite{mcdaniel1973mobility} to model the drift process statistically, calculating a reduced mobility for each drifting species. Finally the overall effective mobility, accounting for distribution of clusters, can be derived by taking a weighted average of true mobilities over cluster populations.

Although this method is not accurate at the sub-percent accuracy of calculations that have been made for the single atomic ion system \cite{McGuirk2009}, the use of DFT allows for the efficient treatment of large molecular ions, which are relatively complex objects, and for which the coupled cluster theory described therein would be prohibitive. The classical parametrization from \cite{mcdaniel1973mobility} is also accurate enough for our purposes, reproducing mobility predictions for bare atomic ions at the few-percent level and all measured data for the Ba$^{+}$ system within its experimental uncertainty.  This validation against previous theoretical work on bare ion mobilities and experimental data for the Ba$^+$ offers a quantification of the accuracy of this method, before extending it to the drift of larger clusters in the Ba$^{++}$ / Xe system.

The remainder of this paper is structured as follows.  Section \ref{sec:Methodology} describes our methodology.  First we describe the calculation of potential surfaces for ion-neutral interactions, then the derivation of the expected cluster population distributions, and finally the combination of this information into ionic mobility predictions.  In Section \ref{sec:Results} we present our main results, including comparisons to data for the Ba$^+$ system and the first predictions for Ba$^{++}$.  Finally in Section \ref{sec:Discussion} we discuss some implications of our results for barium ion transport and tagging in high pressure xenon gas TPC experiments.

\section{Methodology \label{sec:Methodology}}

\subsection{Potential Surfaces and Cluster Energies \label{sec:EdanCalcs}}

\begin{figure}[t]
\begin{center}
\includegraphics[width=0.99\columnwidth]{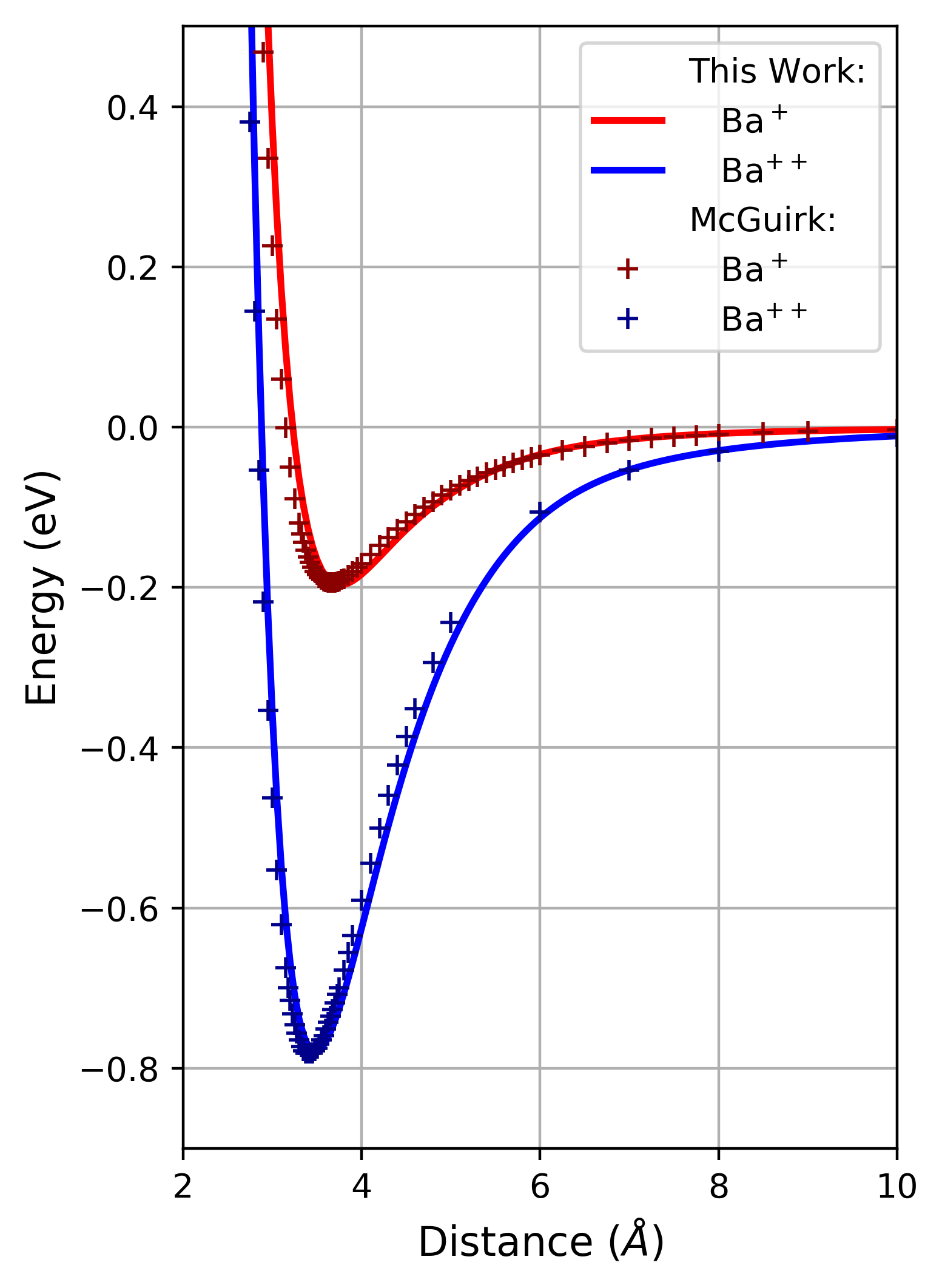}
\caption{Our calculated ion-neutral potential curves from DFT (solid) compared to the coupled-cluster calculations from \cite{McGuirk2009} (dashed) \label{fig:Potentials}}
\end{center}
\end{figure}

\begin{figure*}[t]
\begin{center}
\includegraphics[width=0.99\linewidth]{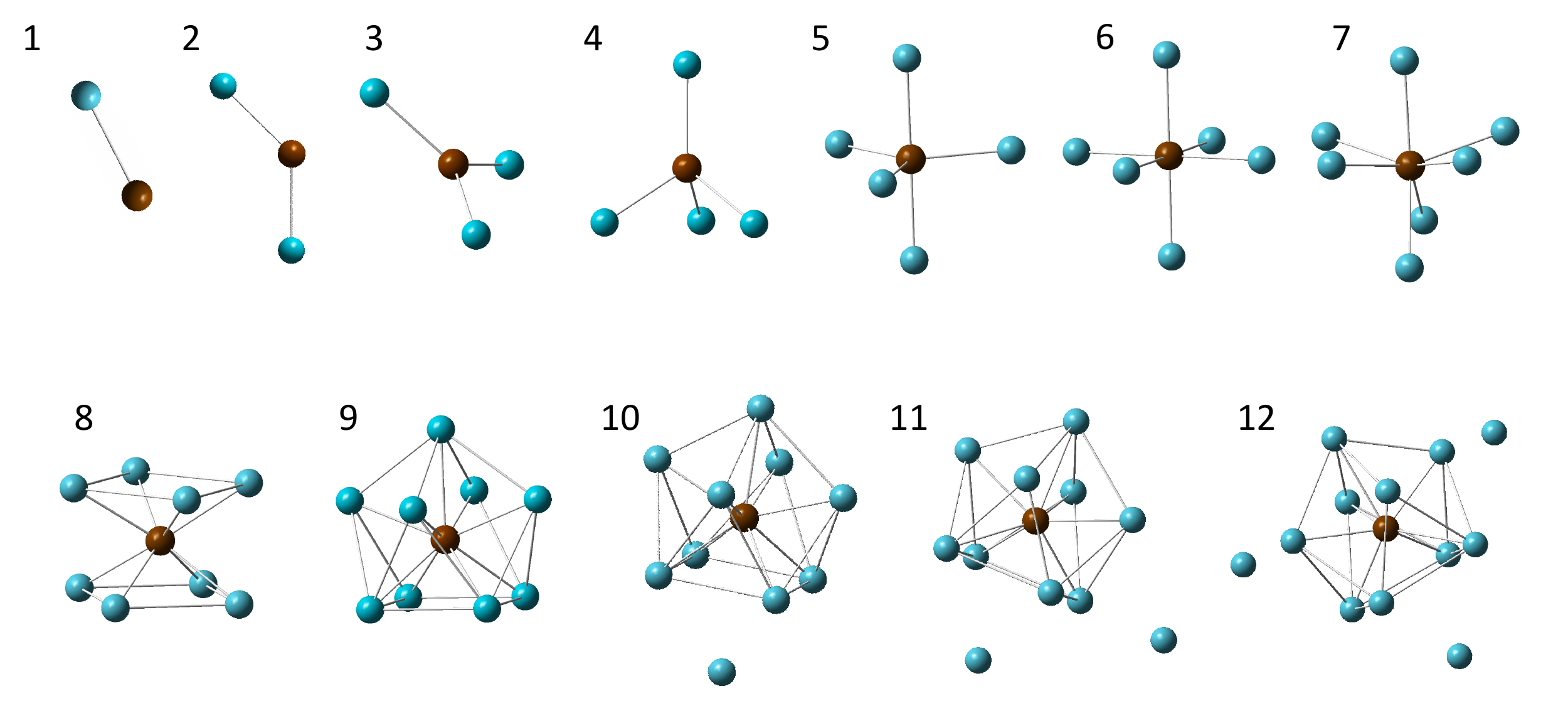}
\caption{Equilibrium conformations of simulated clusters in the Ba$^{++}$ system. \label{fig:BaPPClusterPics}}
\end{center}
\end{figure*}

Potential energy surfaces (PES) and cluster energies are obtained from DFT calculations implemented in the GAUSSIAN '09 package \cite{Frisch2009} using the recommended full Heyd-Scuseria-Ernzerhof hybrid functional (HSE06) \cite{Scuseria2006}. For the expansion of the molecular orbitals, we perform modifications on an initial basis set constructed of the LANL2DZ effective core potential (ECP) \cite{LANL2DZ(Sc-Hg),LANL2DZ(K-Au),LANL2DZ(Na-Bi)}, which replaces core electrons up to Kr for both Ba and Xe with an effective potential \cite{ECP2011}, and a split valence model for the remaining electrons \cite{SVP}. Modifications are strictly made in the valence model.

The basis sets employed in Ref.~\cite{McGuirk2009} yield 5.186 eV and 9.958 eV, respectively, for Ba 1$^{st}$ and 2$^{nd}$ ionization potentials. The 1$^{st}$ ionization potential for Xe was not reported in the paper. Utilizing the same basis set for Xe as Ref.~\cite{McGuirk2009}, we independently calculate it at 12.381 eV. These values reflect errors of 0.50\%, 0.45\%, and 2.07\%, respectively, to experimental data \cite{CRC2004}. Though these are in agreement with the literature and converge well for Ba$^{q+}$ ion-neutral potentials (Fig.~\ref{fig:Potentials}), the bases utilized in Ref.~\cite{McGuirk2009} are too computationally expensive to handle interactions of complex clusters in noble gases. Thus, with the expectation that Ba$^{++}$ will form rather large clusters in Xe gas, the LANL2DZ ECP is chosen for its relatively moderate size to allow for such potential computations while maintaining good agreement with experimental benchmarks (see below).

We carry out the basis set modification in a similar fashion as Ref.~\cite{McGuirk2009}. \textit{s} \& \textit{p} orbitals are taken from the original split valence model. To better account for ion-neutral interactions, polarization functions - angular momentum orbitals of higher \textit{l} number than found in the given species - and diffuse (small exponent) functions are added to the basis set in the form of uncontracted primitive Gaussian functions, $\varphi_\alpha(r) = exp[-\beta_\alpha r^2]$ (Ba: $\beta_p$=0.009, $\beta_d$=0.16, \textbf{0.0355}, 0.021, $\beta_f$=0.5, 0.06; Xe: $\beta_d$=\textbf{0.5805}, \textbf{0.1800}, 0.095, $\beta_f$=2.00,0.40), where bold exponents are taken from the original model. These functions add flexibility and allow for asymmetry in the electronic distributions about atomic centers, as well as extend the tails of atomic orbitals to better simulate atomic charges. Ionization potentials (\%err to exp.) for Ba (1$^{st}$ \& 2$^{nd}$) and Xe (1$^{st}$) are calculated at 5.205 eV (0.14\%), 10.041 eV (0.38\%), and 12.397 eV (1.38\%), respectively. For both Ba$^{+}$ and Ba$^{++}$, we perform full counterpoise corrected \cite{CP1970} PES scans over a sufficiently wide range of Ba$^{q+}$-Xe separation distances as prescribed by Sec.~\ref{sec:IonMobil}. Calculated values at this level of theory match well with the behavior observed by Ref.~\cite{McGuirk2009}, as shown in Fig.~\ref{fig:Potentials}. 

With the level of theory established and verified, we relax [BaXe$_N$]$^{q+}$ clusters ($q$=1, $N$=1-4; $q$=2, $N$=1-12) to respective global minima (symmetry consideration turned off).  The equilibrium conformations for the $q$=2 structures are shown, for illustration, in Fig.~\ref{fig:BaPPClusterPics}. Note that in clusters of $N>9$, additional Xe atoms lie in a more distant outer shell with far weaker binding. After relaxation we perform vibrational frequency analyses \cite{Thermo2000,Vibration1999} at STP to obtain cluster enthalpies and entropies. These values are used to determine the distribution of clusters as described in Sec.~\ref{sec:EqmConst}. Lastly, we perform PES scans for the [BaXe$_N$]$^{q+}$ clusters. For each cluster, we define a center-of-mass coordinate system and sweep through a variable set of ($\phi,\theta$) coordinates at 15$^{\circ}$ to 30$^{\circ}$ intervals, where $\phi$ and $\theta$ are the azimuthal and polar angles, respectively. These curves are used as prescribed in Sec.~\ref{sec:IonMobil}. to derive ion mobilities. 

The potential curves are augmented at large distances by a linear regression logarithmic fit of the scan performed at the Hartree-Fock level of theory, which provides a good match to the shape of the potential in the cases where such large-distance behaviour could not be calculated with adequate convergence.

\subsection{Cluster Population Distributions \label{sec:EqmConst}}

Cluster population distributions are derived from the equilibrium constants $K^q_N$ for the reactions described by:
\begin{equation}
    \mathrm{[BaXe_N]}^{q+}\leftrightarrow \mathrm{[Ba_{N-1}]^{q+}} + \mathrm{Xe} \label{eq:EqmConsts}
\end{equation}
The equilibrium constant for gases is defined in terms of the activity of each component via:
\begin{equation}
    K^Q_N=\frac{a_\mathrm{[BaXe_N^{q+}]}}{a_\mathrm{[Ba_{N-1}^{q+}]}a_\mathrm{Xe}}
\end{equation}
where for ideal gases the activity is defined as the ratio of the partial pressure $p_N$ at equilibrium to standard pressure $p_0$ = 1 bar: $a_N=p_N/p_0$. The equilibrium constant for each reaction is predictable based on the standard reaction entropy and enthalpy via the Van 't Hoff equation \cite{atkins2014atkins}:
\begin{equation}
    K=Exp\left[-\frac{\Delta H_\ominus}{RT}+\frac{\Delta S_\ominus}{T}\right].
\end{equation}
For the reactions in question we can predict the entropy and enthalpy changes based on the calculations of Sec.~\ref{sec:EdanCalcs}.  These are used to establish the equilibrium constants for each step, from which the relative populations of each species $\mathrm{[BaXe_N^{q+}]}$ can be derived iteratively.

\subsection{Ionic Mobility Calculations \label{sec:IonMobil}}

The mobility of each ionic species is calculated via the low-field ion mobility equation \cite{mcdaniel1973mobility}:
\begin{equation}
\mu_{0}(0)=\frac{\zeta}{3^{1/2}}\left(\frac{1}{m}+\frac{1}{M}\right)^{1/2}\frac{q}{(k_B T)^{1/2}\sigma_Q}\frac{1}{N}.
\label{eq:MobilityEq}\end{equation}
Here, $\zeta$ is a numerical factor which can be calculated to be 0.814 from Chapman Enskog theory \cite{chapman1970mathematical,hirschfelder2003molecular}. $m$ and $M$ are the masses of the two colliding objects.  To facilitate comparisons with experimental data and previous theoretical work, we assume the masses of the most prevalent naturally occurring isotope of barium $^{137}$Ba and the isotopic average mass of natural xenon.  The alternative case of interest of double beta decay experiments, $^{136}$Ba in isotopically enriched $^{136}$Xe is discussed in Section \ref{sec:Discussion}; q is the total charge; $k_B T$ is the thermal energy, set here to 300K \footnote{Note that we evaluate all mobilities at 300K, but following convention we report the reduced mobility, defined as $K_0=\frac{T_0}{T}K$ with T$_0$=273K};  $\sigma_Q$ is the momentum transfer cross section; and N is the total number density of scatterers.

The momentum transfer cross section Q is derived from the shape of the potential surface for the cluster in question. Following the prescription of \cite{mcdaniel1973mobility} we make the first-order classical approximation to the collision integral,
\begin{equation}
\sigma_{Q}=2\pi\int_{0}^{\infty}(1-cos\theta)b\,db,
\end{equation}
where b is the impact parameter and $\theta$ is given by:
\begin{equation}
\theta=\pi-2b\int_{r_{a}}^{\infty}\left[1-\frac{b^{2}}{r^{2}}-\frac{V(r)}{E}\right]^{-1/2}\frac{dr}{r^{2}},
\end{equation}
with $r_{a}$ the classical distance of closest approach. $Q$ is averaged over the thermal collision spectrum to yield the first order collision integral in steps of 0.02$\times kT$ according to the recipe in  \cite{mcdaniel1973mobility}.  The second order correction to the collision integral was also evaluated, but found to be negligible in all cases.

For the non-spherical structures we also make the Mason Monchick approximation \cite{monchick1961transport,viehland2001ion}, taking the effective cross section to be the average of pseudo-spherical potentials approached from different directions, weighted by solid angle. 

Using the above ingredients we calculate the reduced mobility of each molecular ion species.  Under conditions when several molecular ions are present in significant populations, the mobility $\mu^0_{eff}$ will be the partial-pressure-weighted average of their respective mobilities $\mu^0_N$:
\begin{equation}
    \mu^0_{eff}=\sum_N \left[p_N \mu^0_{N}\right] / \sum_N p_N.
\end{equation}

In this way, the mobility of ions in equilibrium with xenon gas can be predicted as a function of xenon pressure and temperature.

\section{Results \label{sec:Results}}

\begin{figure}[b]
\begin{center}
\includegraphics[width=0.9\columnwidth]{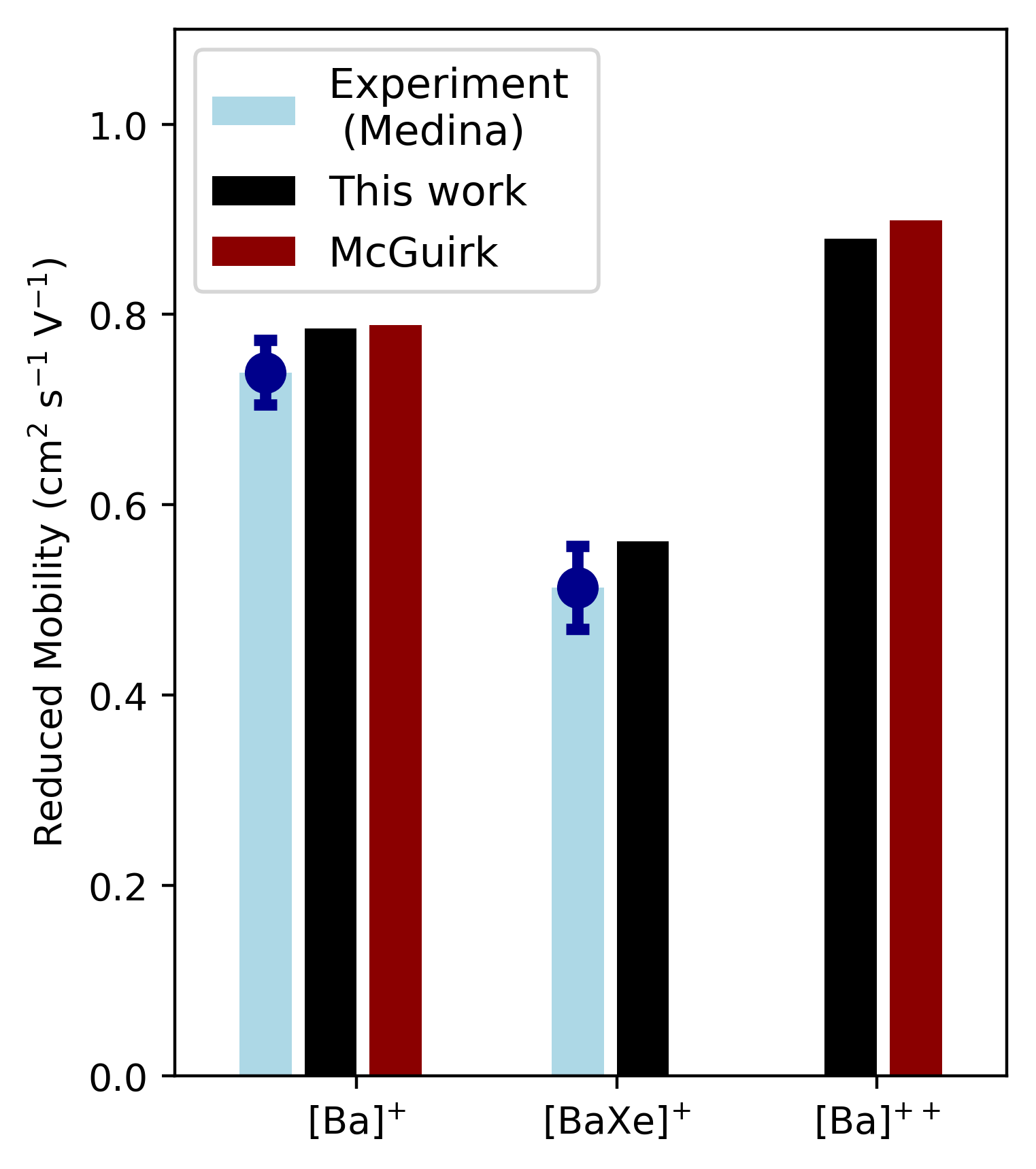}
\caption{Validation of our method: reduced mobilities of Ba$^+$, [BaXe]$^+$ and Ba$^{++}$ compared with values extracted from experimental data \cite{Cesar2014} and past theoretical work. \cite{Viehland2017}.\label{fig:MobAndEqmBaPlus}}
\end{center}
\end{figure}

Here we present the results of the calculations described above. We first benchmark our technique using the Ba$^{+}$ system and then make the first predictions for the Ba$^{++}$ system.

\subsection{The Ba$^{+}$ system}

\begin{figure}[t]
\begin{center}
\includegraphics[width=0.99\columnwidth]{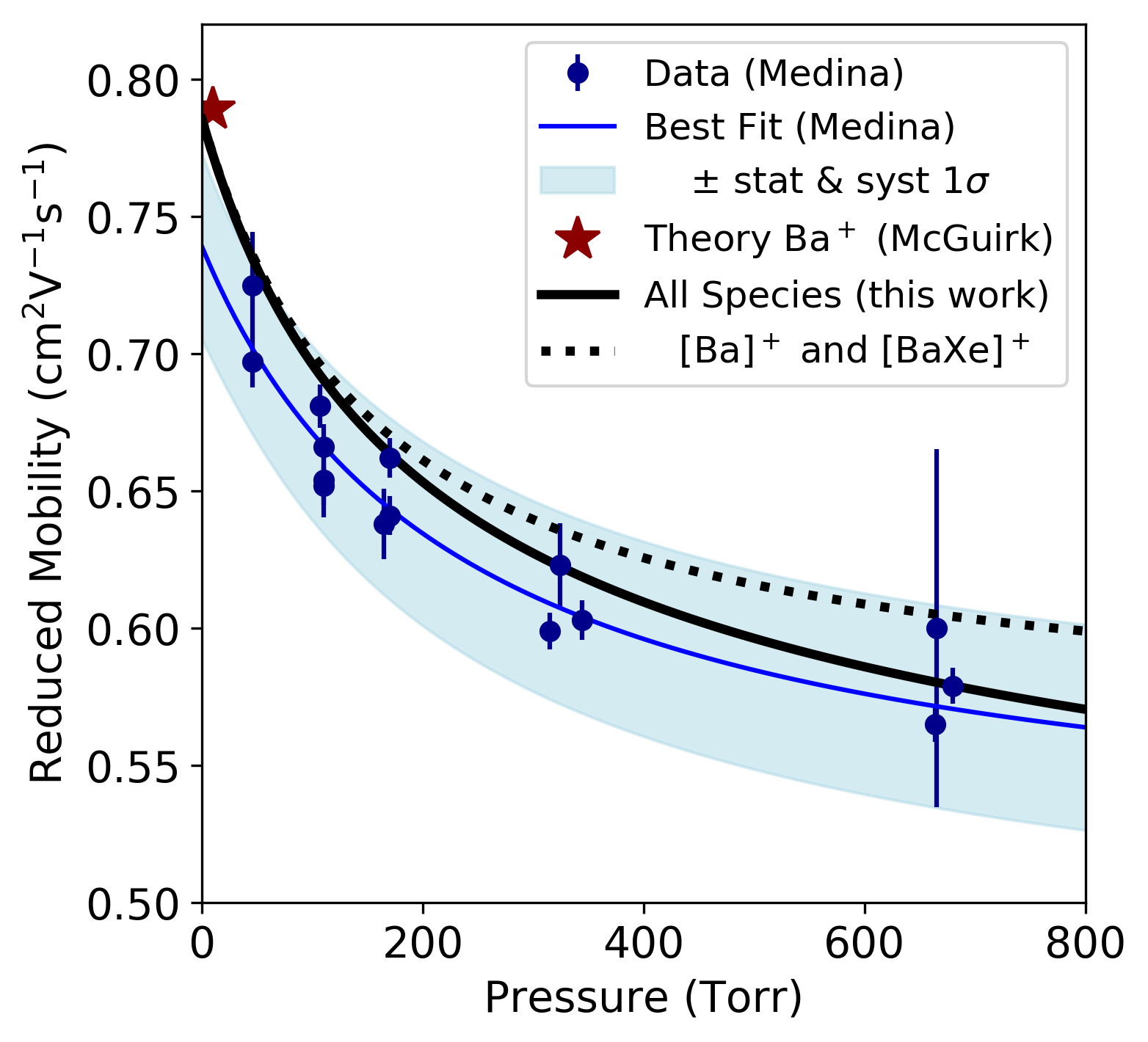}
\caption{Comparison of our predicted effective reduced mobility to experimental data from~\cite{Cesar2014}.  Equilibrium constants are evaluated at 296K as specified in~\cite{Cesar2014}.  \label{fig:MedinaComparison}}
\end{center}
\end{figure}

Our first goal for the Ba$^{+}$ system is comparison with data from Ref.~\cite{Cesar2014}, which requires a prediction of mobilities in the 0-1 bar range.  The pressure dependent mobility observed in that work was fitted to an empirical function which accounted for formation of [BaXe]$^{+}$ molecular ions as a function of pressure, and the mobility of the [Ba]$^{+}$ and  [BaXe]$^{+}$ and the dissociation constant were extracted, with accompanying statistical and systematic uncertainties.  The [Ba]$^{+}$ mobility was found to be consistent with, but slightly lower than the precise theoretical predictions of \cite{Viehland2017}.  

Fig.~\ref{fig:MobAndEqmBaPlus}, shows our predictions for the mobilities of [Ba]$^{+}$, [BaXe]$^{+}$ and [Ba]$^{++}$ compared with the values extracted from experiment,  and with theoretical predictions as calculated in \cite{McGuirk2009} and updated in \cite{Viehland2017}. We find good agreement with the past theoretical work, slightly over-predicting the reduced mobilities extracted from data, but consistent with them at the 1$\sigma$ level.  We also predict the relative molecular ion populations.  We find that the in 0-1 bar pressure range, in addition to the two species assumed in simple model of \cite{Cesar2014}, at the highest pressures there are also sizeable populations of [BaXe$_2$]$^{+}$ and [BaXe$_3$]$^{+}$ that contribute to the mobility.  The predicted population distributions are shown in Fig.~\ref{fig:ClusterPopulations}, top. We can calculate the pressure dependent mobility in the 0-1 bar range assuming either a simple two-species equilibrium as in Ref~\cite{Cesar2014}, or alternatively including all relevant ionic species with the ratios shown in Fig.~\ref{fig:ClusterPopulations} (our most accurate prediction).  Both models are shown compared to data and the empirical fit in Fig.~\ref{fig:MedinaComparison}.  It is clearly observed that inclusion of the higher mass clusters improves agreement with experimental data at higher pressures.  This explains, to some extent, the mild tension observed in Fig.~\ref{fig:MobAndEqmBaPlus}, where the values from experiment implicilty assume a two-species model for their definition.

With our full first-principles model we thus find strong agreement with experimental data in terms of both absolute mobility and pressure dependence in the Ba$^{+}$ system.  This validation gives confidence in our predictions of both mobilities and equilibrium constants.  We extend our pressure dependend mobility predictions for the Ba$^{+}$ system to higher pressures in Fig.~\ref{fig:MobilityVsPressure}, and proceed to the more complicated doubly charged system, for which no prior data or theoretical calculations exist.

\subsection{Ba$^{++}$ system}

\begin{figure}
\begin{center}
\includegraphics[width=0.99\columnwidth]{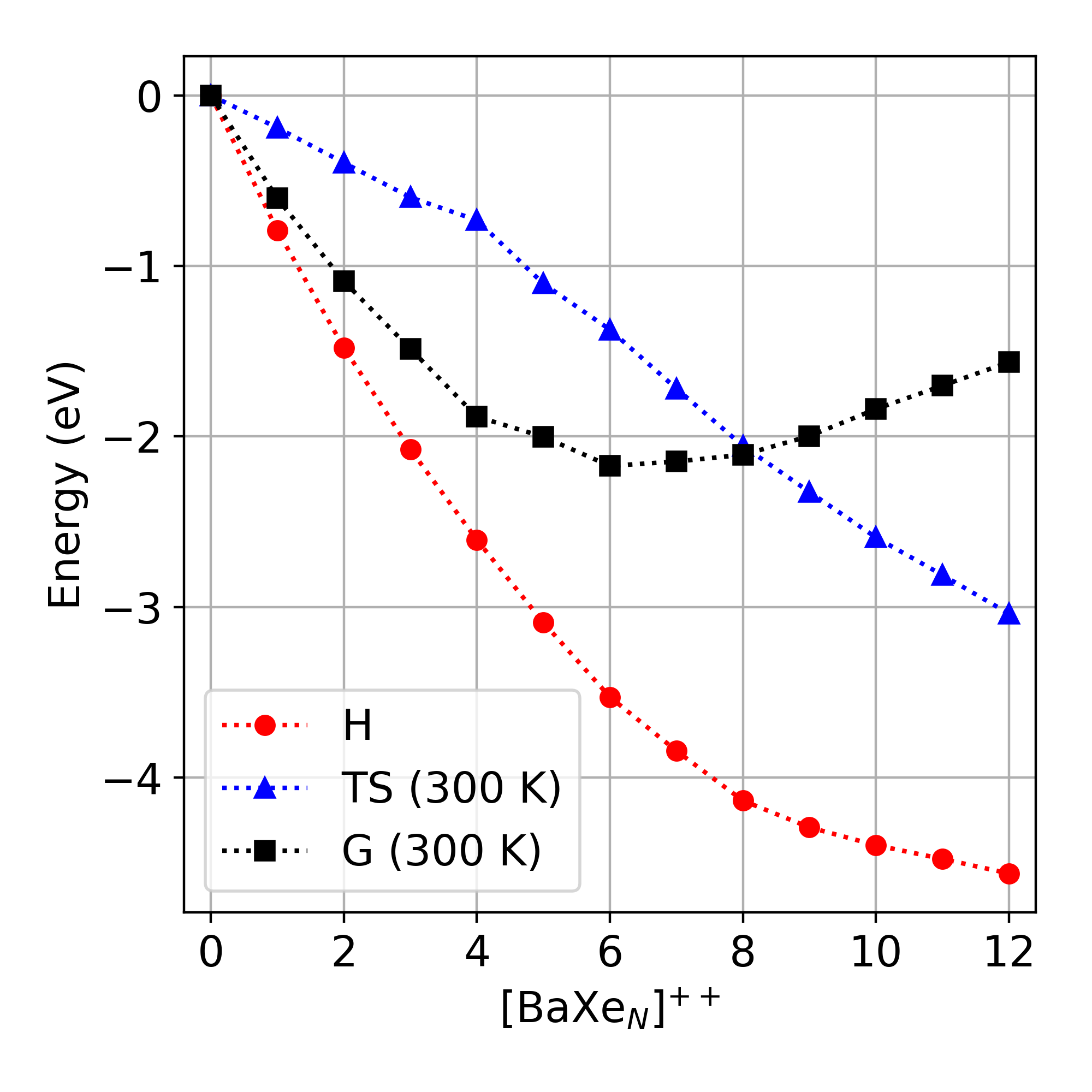}

\caption{Standard enthalpy, entropy and Gibbs energy of formation for clusters in the Ba$^{++}$ system at 300K.  \label{fig:EnthEntBaPP}}
\end{center}
\end{figure}

In the Ba$^{++}$ system, the potential well is substantially deeper, which allows the formation of larger and more complex clusters.   Calculated equilibrium configurations for all clusters up to N=12 are shown in Fig.~\ref{fig:BaPPClusterPics}.   We observe that significant enthalpy changes are observed for clusters of size up to N=10 (Fig.~\ref{fig:EnthEntBaPP}). 

\begin{figure}[t]
\begin{center}
\includegraphics[width=0.99\columnwidth]{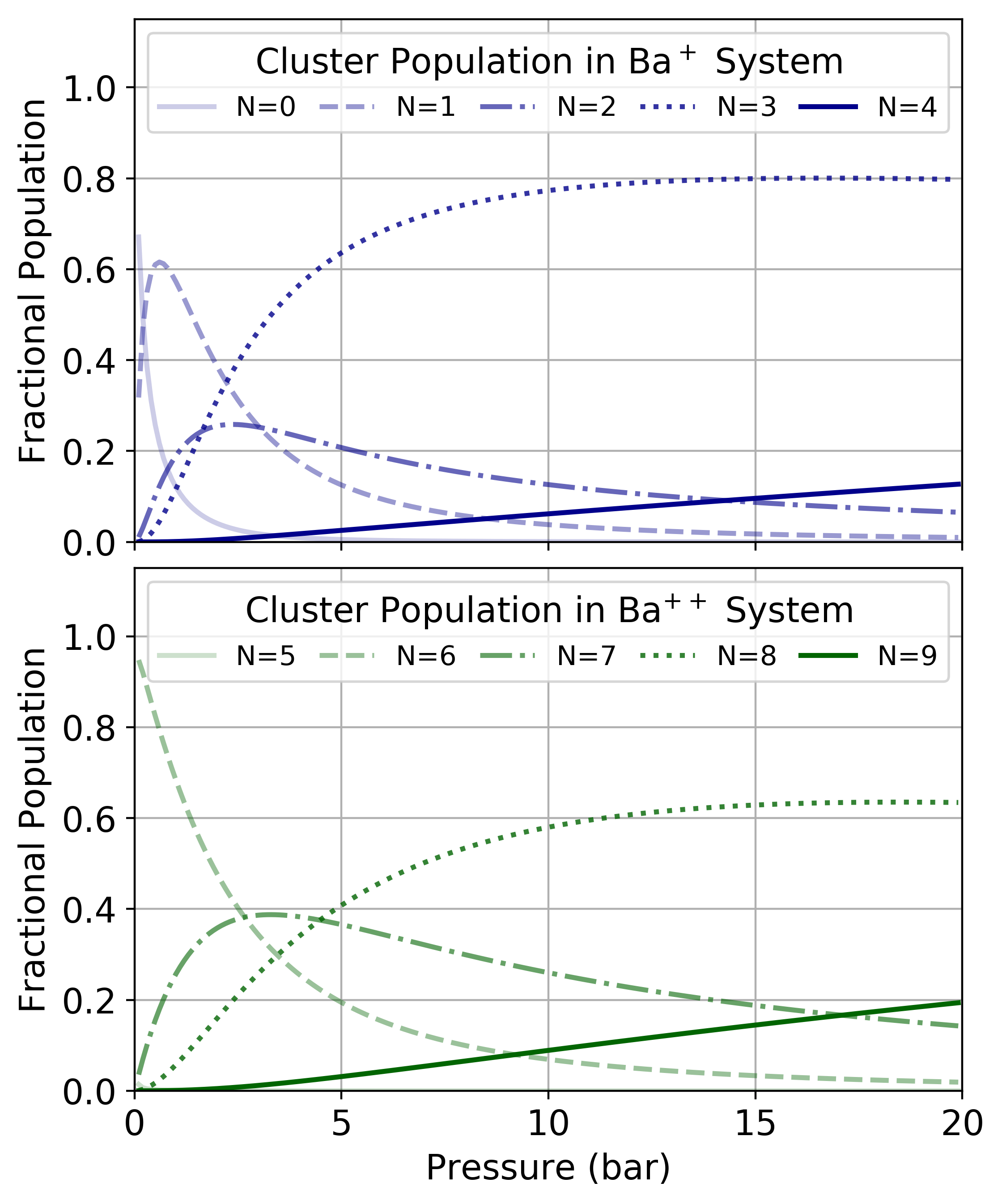}

\caption{Cluster populations from thermodynamic calculations \label{fig:ClusterPopulations}}
\end{center}
\end{figure}

The exponent of the equilibrium constant for each reaction is plotted in Fig.~\ref{fig:EnthEntBaPP}, along with the values of $\Delta S_\ominus$ and $\Delta H_\ominus$.  The shape of the Gibbs energy curve shows that forward reactions are very favorable up to N=6, and reverse reactions are very favorable above N=6, at STP.  However, because Gibbs energy varies rather smoothly around N=6, the population of different species will be a strong function of pressure.  The distribution of ion populations in our pressure range of interest is shown in Fig.~~\ref{fig:ClusterPopulations}, bottom.  We observe that under conditions of interest, the mobilities of clusters 5$\leq$N$\leq$9  are experimentally relevant.

\begin{figure}[t]
\begin{center}
\includegraphics[width=0.99\columnwidth]{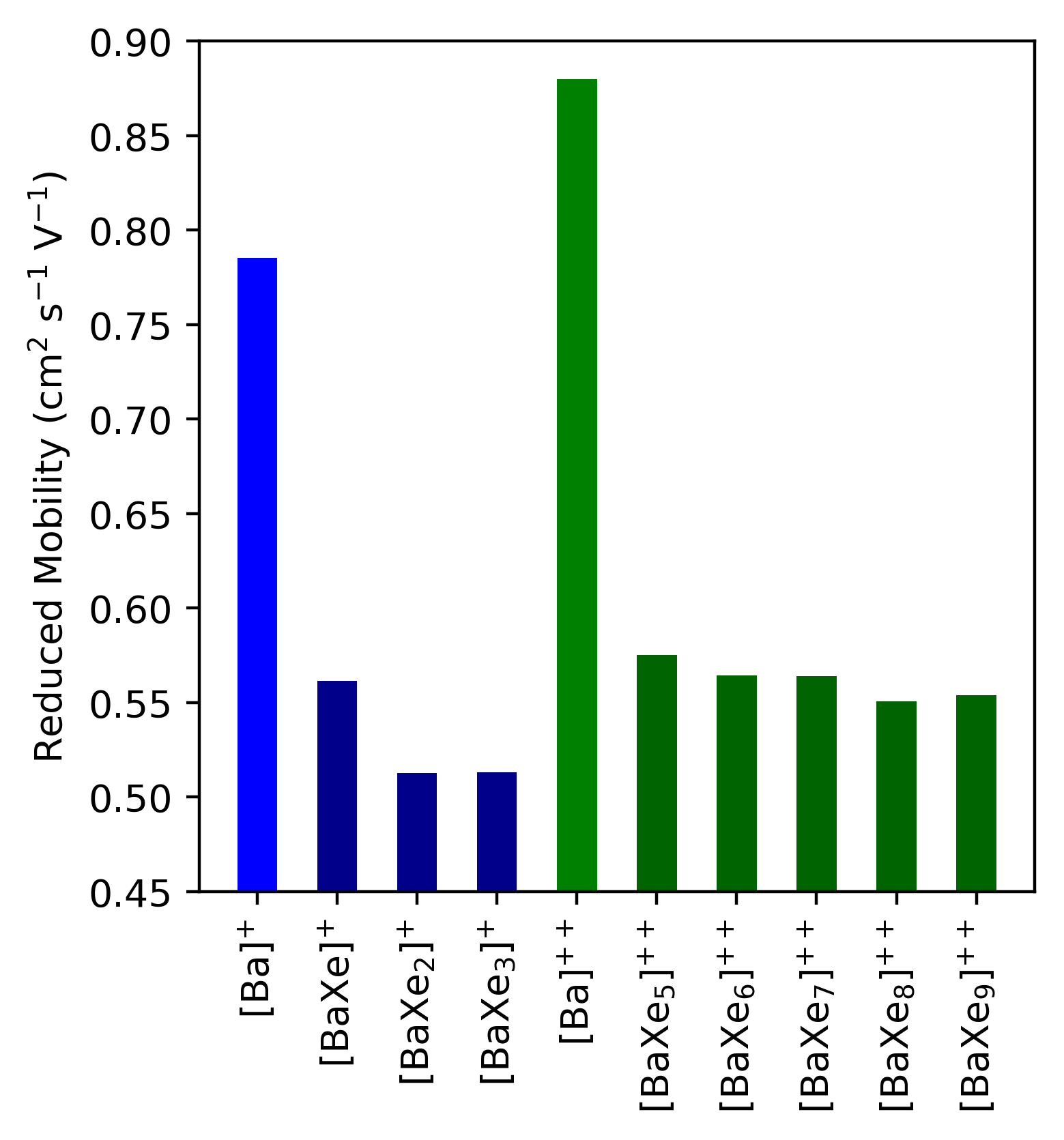}

\caption{Reduced mobilities of all clusters considered in this work. \label{fig:AllClusters}}
\end{center}
\end{figure}

The reduced mobilitites are calculated for these clusters and shown in Fig~\ref{fig:AllClusters}.  The average mobility is the population-weighted average of these quantities and is shown in Fig.~\ref{fig:MobilityVsPressure}.

Because these clusters are all large, their effective radius does not change drastically as a function of N.  Thus the mobilities have fairly soft N dependence, which translates to a very weak pressure dependence of Ba$^{++}$ effective mobility relative to Ba$^{+}$.  The extra unit of charge on Ba$^{++}$ ultimately outweighs the enhanced cluster mass and radius, resulting in a higher mobility for the Ba$^{++}$ species at pressures above 1 bar.

\section{Discussion\label{sec:Discussion}}

We have calculated cluster properties and distributions for [BaXe$_N$]$^{q+}$ molecular ions in xenon gas and studied the mobilities of the dominant species in the Ba$^{+}$ and Ba$^{++}$ systems at 0-20 bar.  Our model is in excellent agreement with experimental data in for Ba$^{+}$ system, accurately predicting the pressure dependent mobility.  The extracted vales of mobility for individual species and equilibrium position are in relatively good agreement with empirical fits to experiment, although these fits assume a model that is less complete in terms of relevant species than this work, somewhat clouding this comparison.  For the Ba$^{++}$ system we find good agreement with past theoretical works for the reduced mobility of the bare atomic ion, and we predict a cluster distribution that peaks around N=6-8 in the pressure range 1-20 bar, with the population and hence reduced mobility having a non-trivial but weak pressure dependence due to the shifting molecular ion equilibrium.

In addition to mobility, some other notable properties of the [BaXe$_N$]$^{++}$ system are of interest from the point of view of barium tagging experiments. A highly relevant question for neutrinoless double beta decay experiments is: is there a sizeable rate constant for spontanous charge transfer from Ba$^{++}$ to Ba$^{+}$ and Xe$^{+}$?  For barium tagging methods that are specific to the dication state, such as in Ref.~\cite{McDonald:2017izm}, a large rate of dissociation would inhibit the sensitivity of the technique.  Ignoring the effects of molecular ion formation, this dissociation would not occur due to the large difference in ionization potentials between barium and xenon.  However, the analysis here presents a somewhat more nuanced picture.  Now, the most favorable neutralization reactions in the presence of the xenon shell will processes such as:
\begin{equation}
\mathrm{[BaXe_{N}]^{2+}\rightarrow[BaXe]^{+}+[XeXe]^{+}+\left(N-3\right)Xe}\label{eq:Neutralization}
\end{equation}
We can calculate the rate constant for this reaction for any N using the calculated standard entropies and enthalpies evaluated above.  We find that the reaction is always highly disfavored with equilibrium constants of 10$^{-38}$ or smaller for all relevant clusters.  It thus appears highly unlikely that spontaneous neutralization will occur, even in the presence of clustering, and the doubly charged state will remain stable in bulk xenon.

The total binding enthalpy of the cluster is also of interest, since this can in principle inhibit chemical reactions involving the barium dication. Such reactions are critical to some barium tagging methods  \cite{Jones:2016qiq}.  Based on our calculations, the total binding enthalpies of relevant clusters are 3-4  eV/ion, which must be supplied in order to remove the barium dication from the cluster. On the other hand, the enthalpy of solvation of the Ba$^{++}$ ion in water is 13.5 eV/ion. Thus removing a barium ion from a xenon shell appears to be substantially more energetically favorable than removing it from a solvation shell of water, and any reactions which proceed in water might thus be expected to proceed in high pressure xenon gas.  

\begin{figure}[t]
\begin{center}
\includegraphics[width=0.98\columnwidth]{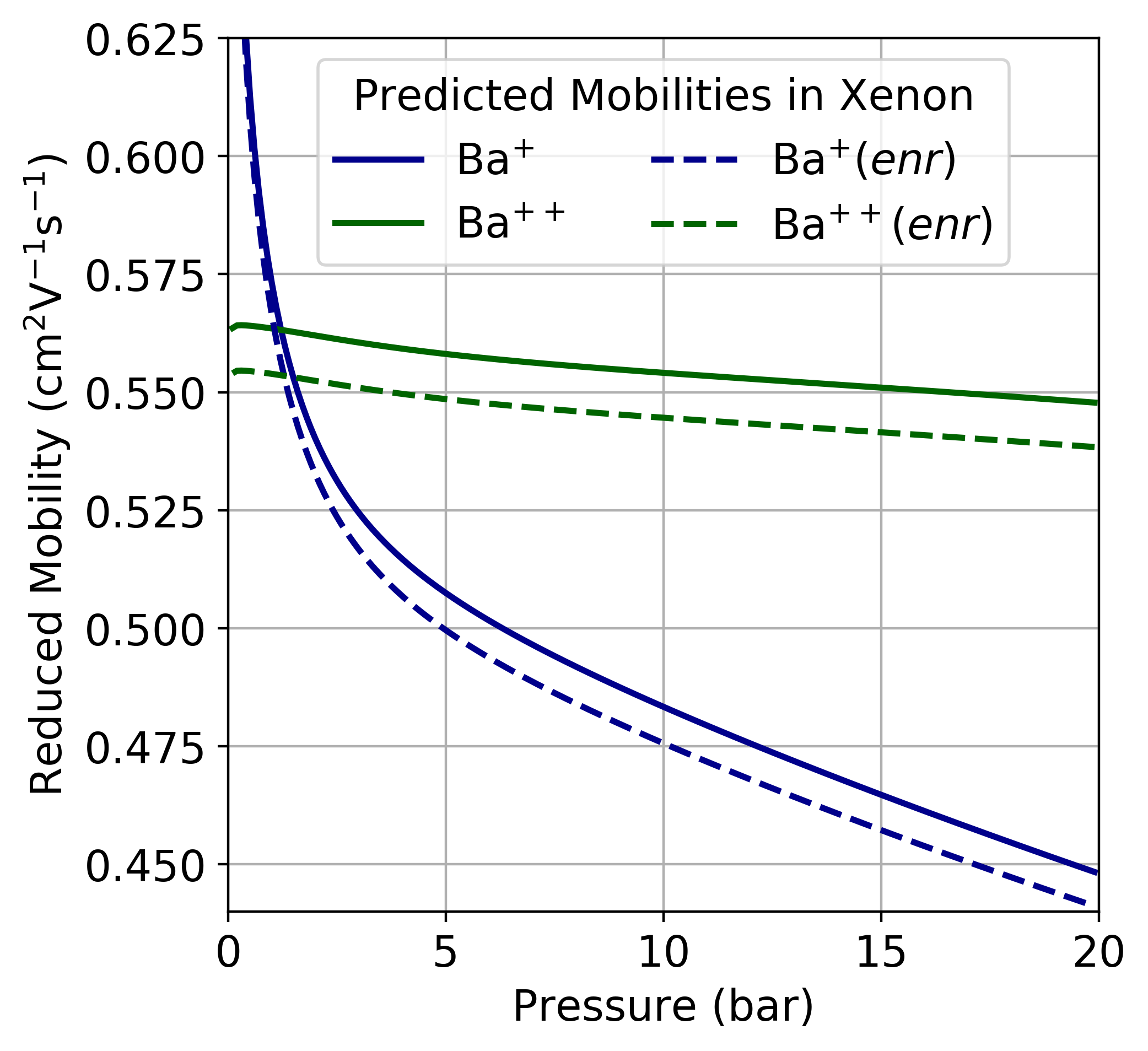}

\caption{Our calculated pressure dependent mobility for the Ba$^{+}$ and Ba$^{++}$ systems accounting for the effects of molecular ion formation. \label{fig:MobilityVsPressure}}
\end{center}
\end{figure}

It is also notable that in future high pressure xenon TPCs for neutrinoless double beta decay, the operating medium will be isotopically enriched $^{136}$Xe and the drifting isotope of barium will be exclusively $^{136}$Ba. This has a small impact on the mobilities relative to those shown here, which are evaluated using the most prevalent naturally occuring isotope of barium and the average atomic mass for xenon.  This slight mobility reduction can be calculated by substituting the mass-dependent term in Eq.~\ref{eq:MobilityEq}. For all species considered here this effect leads to a 0.5-1.5\% level reduction in drift velocity.  The mobility of pure $^{136}$Ba in pure $^{136}$Xe is shown as dashed curves in Fig.~\ref{fig:MobilityVsPressure}

Comparison with both data and previous theoretical works suggest that the analysis presented here to be accurate at the few-percent level. However, experimental verification is mandatory.  We expect to confront these predictions with data from our laboratory in the near future, as part of the R\&D program to develop barium tagging techniques for high pressure xenon gas time projection chambers.

\acknowledgments

This work was supported by the University of Texas at Arlington. BJPJ and DRN are supported by the Department of Energy under grant numbers DE-SC0017721.  E. Bainglass and M. N. Huda are supported by NSF award no. 1609811.  We acknowledge the NEXT collaboration for their input on this work.

\bibliographystyle{apsrev4-1}
\bibliography{main}

\end{document}